\documentstyle[12pt]{article}

\def\beq{\begin{equation}} \def\eeq{\end{equation}}
\def\bea{\begin{eqnarray}} \def\eea{\end{eqnarray}}
\let\nn=\nonumber
\def\beann{\begin{eqnarray*}} \def\eeann{\end{eqnarray*}}

\let\a=\alpha \let\be=\beta \let\g=\gamma \let\de=\delta
\let\e=\varepsilon   
  \let\la=\lambda 
  \let\p=\pi  \let\s=\sigma

\let\qd=\quad

\def\0{\over } \def\1{\vec }     \def\2{{1\over2}} \def\4{{1\over4}}
\def\5{\bar }  \def\6{\partial } \def\7#1{{#1}\llap{/}}

\def\<{\langle } \def\>{\rangle }

\def\i{{\rm i}}

\def\ctg{\mbox{ctg}}

\begin{document}
\markright{revised version, 5/7/95, cond-mat/9506077}

\begin{center}
{\Large {\bf A Note on Inverse-Square Exchange Models\\}}
\vspace{2cm}
{\large Frank G\"{o}hmann
\footnote[2]{e-mail: frank@monet.phys.s.u-tokyo.ac.jp}
and Miki Wadati}\\
\vspace{5mm}
Department of Physics, Faculty of Science, University of Tokyo,
Hongo 7-3-1, Bunkyo-ku, Tokyo 113, Japan\\
\vspace{15mm}

{\large {\bf Abstract}}
\end{center}
\begin{list}{}{\addtolength{\rightmargin}{10mm}
               \addtolength{\topsep}{-5mm}}
\item
The su(1$|$1) symmetric version of the Haldane-Shastry spin chain is
diagonalized by means of a linear transformation. The same
transformation applied to the original su(2) model yields simple
expressions for the Hamiltonian and the generators of the Yangian
symmetry of the model in terms of spin wave operators.
\end{list}

\vspace{2cm}

Several years ago Haldane and Shastry independently discovered a soluble
spin chain with exchange interaction falling off as the inverse square
of the distance between the spins \cite{Haldane88,Shastry88}. This
model can be thought of as the static limit \cite{Polychronakos93} of
the Sutherland model with internal degrees of freedom
\cite{HaHa92,HiWa93a,KaKu95}. The idea of putting internal degrees of
freedom into the Calogero-Sutherland models has proved to be very
fruitful since then. A rich family of related models has been found.
These long range interacting systems are under active investigation
now. A common algebraic structure underlying their integrability is
still waiting for its discovery.

In this letter we discuss a linear transformation, that turns
the su(m$|$n) supersymmetric extension of the original
Haldane-Shastry Hamiltonian into a more compact form, consisting
of a sum over the elementary excitation energies times some
operator. For brevity we focus on the su(1$|$1) an su(2)
versions of the spin chain. We find that our transformation is
the explicit diagonalizing transformation for the su(1$|$1)
model, which has been known by numerical investigation to be
equivalent to a free fermion model \cite{Haldane94}. The su(2)
model is not diagonalized by that transformation. But the
Hamiltonian takes on a compact form, which allows, for example,
for an easy algebraic derivation of the ferromagnetic spin wave
spectrum.

The most obvious generalizations of the original Haldane-Shastry
su(2) $N$-particle system
\beq \label{1}
     H = \sum^{N-1}_{j,k = 0 \atop j < k}
         \sin^{-2} ((j - k) \p /N)(P_{jk} - 1)
\eeq
are obtained by replacing the su(2) representation of the pair
permutation operator $P_{jk}$ by su(n) representations or by
supersymmetric su(m$|$n) representations. As has been stated
by Haldane \cite{Haldane94}, in the su(1$|$1) case, when
$P_{jk}$ is expressed by Fermi operators as
\bea \label{2}
     P_{jk} & = & 1 - (c^+_j - c^+_k)(c_j - c_k) \qd,
     \\ \label{3}
     \{ c_j^+ , c_k \} & = & \de_{jk} \qd, \qd \{ c_j , c_k \} =
     \{ c_j^+ , c_k^+ \} = 0 \qd,
\eea
the model is equivalent to a free fermion model.

The linear transformation that diagonalizes $H$ in the representation
(\ref{2}) is nothing but a discrete Fourier transform.
There is the following remarkable discrete Fourier
representation of the cotangent:
\beq \label{4}
     1 - \i \, \ctg (j\p /N) = \frac{1}{N} \sum_{k=0}^{N-1}
                            (N - 2k) \exp( - \i 2 \p jk/N) \qd,
\eeq
for $j = 1, \dots , N - 1$. This formula follows from the geometric sum
formula. Since
\beq \label{5}
     \sin^{-2} (x) = (1 + \i \, \ctg (x))(1 - \i \, \ctg (x))
\eeq
we get with (\ref{4}) and (\ref{5}) in (\ref{1})
\bea \label{6} \nn
     H & = & - \frac{1}{N} \sum_{j,k = 0 \atop j < k}^{N-1}
               \sum_{m,n = 0}^{N-1} (N - 2m)(N - 2n)
               \exp( \i 2\p (j - k)(m - n)/N)
               (c^+_j - c^+_k)(c_j - c_k)  \\ \label{7}
       & = & 2 \sum_{k=1}^{N-1} k(k - N)
               \tilde{c}_k^+ \tilde{c}_k
\eea
where
\beq \label{8}
     \tilde{c}_k := \frac{1}{\sqrt{N}} \sum_{j=0}^{N-1}
                    \exp( - \i 2\p jk/N) c_j
\eeq
and $\tilde{c}_k^+ = (\tilde{c}_k)^+$. The step from the first
to the second line in (\ref{7}) requires only elementary calculations.
Since the operators $\tilde{c}_k$ are again Fermi operators with
commutation relations (\ref{3}), $H$ as expressed in (\ref{7}) is
in fact diagonal. The densities $n_k := \tilde{c}_k^+
\tilde{c}_k$ form a set of commuting conserved quantities, and,
in principle, there is nothing left to say about the model.

Let us however comment on one of its possible origins. The su(1$|$1)
Hamiltonian (\ref{1}) appears as the static limit of a
''minimal`` supersymmetric extension of the Sutherland model: Given
the ground state $\psi := \exp(\chi)$ with energy $E_0$ of any bosonic
Hamiltonian $H_b$ with kinetic energy equal to $- \sum_l \6_l^2$,
this Hamiltonian may be factorized as
\bea
     \tilde{H}_b & := & 2(H_b - E_0) = \sum_l q_l^+ q_l \qd, \\
             q_l & := & (\6_l \chi) - \6_l \qd.
\eea
With the definition
\beq
     Q := \sum_l q_l c_l^+ \qd,
\eeq
$c_l^+$ being Fermi operators according to (\ref{3}), a ''minimal``
supersymmetric extension off $H_b$ is \cite{ABEI85}
\beq \label{b4}
     \hat{H} = \tilde{H}_b + \sum_{l,m} [ q_l , q_m^+ ] c_l^+ c_m \qd.
\eeq
By construction the operators $\hat{H}, Q^+, Q$ fulfil the
usual supersymmetric algebra,
\beq
     \{Q, Q^+ \} = \hat{H} \qd, \qd \{Q, Q\} = \{Q^+ , Q^+ \} = 0
                   \qd.
\eeq
Hence $Q$ and $Q^+$ commute with $\hat{H}$. Furthermore, the Fermion
number is conserved. Therefore the Hamiltonian $\hat{H}$ is
block diagonal in a Fermion number basis, and the different
blocks may be considered as a supersymmetric hierarchy of
Hamiltonians with related spectra \cite{ABEI85}.

In case of $H_b$ being the Sutherland Hamiltonian, the
supersymmetric operator $\hat{H}$ is especially simple due to the
fact that the ground state factorizes into a pair product.
Inserting the ground state wave function of the Sutherland
Hamiltonian \cite{Sutherland71c} into (\ref{b4}) one gets
\beq
     \hat{H} = - \sum_{j=1}^{N} \6_j^2 + 2 \la
                 \sum_{j,k = 1 \atop j < k}^N
                 \sin^{-2} (x_j - x_k)
                 (\la - (1 - (c_j^+ - c_k^+)(c_j - c_k))) \qd.
\eeq
(cf.\ \cite{ShSu93}). Freezing out the dynamical degrees of freedom
\cite{Polychronakos93} one arrives, upon some rescaling and
renumbering, at the su(1$|$1) version of (\ref{1}).

What is happening, if we apply the transformation (\ref{4}) to
the su(2) representation of the Hamiltonian? In this case
\beq \label{9}
     P_{jk} = \frac{1}{2} \left( 1 + \s_j^{\a} \s_k^{\a} \right)
              - \de_{jk}
\eeq
where, as usual, $\s_j^{\a}$ acts as Pauli matrix $\s^{\a}$ on
the state of the $j$-th spin and as unit operator on the other
spins. Inserting (\ref{4}) and (\ref{9}) into (\ref{1}) yields
\beq \label{10}
     H = 2 \sum_{k=1}^{N-1} k(k - N) \left( \frac{1}{N}
         \tilde{S}_{-k}^{\a} \tilde{S}_{k}^{\a} - \frac{1}{2}
         \right) \qd.
\eeq
The $\tilde{S}_{k}^{\a}$ are spin wave operators,
\beq \label{11}
     \tilde{S}_{k}^{\a} := \frac{1}{2} \sum_{j=0}^{N-1}
                           \exp( \i 2\p jk/N) \, \s_j^{\a} \qd,
\eeq
with commutation relations
\beq \label{12}
     \left[ \tilde{S}_j^{\a} , \tilde{S}_k^{\be} \right] =
        \i \, \e_{\a \be \g} \tilde{S}_{j+k}^{\g} \qd,
\eeq
periodic in $j$ and $k$ with period $N$. Because of the non-local
nature of these commutation relations, the operators
$\tilde{S}_{-k}^{\a} \tilde{S}_k^{\a}/N - 1/2$ do not commute
among each other for different values of $k$, and (\ref{10}) is
not a diagonal Hamiltonian. Furthermore, the Heisenberg equations
of motion for the spins remain non-local.

The representation (\ref{10}) of the Hamiltonian is of its own
beauty, and may prove to be useful for further investigations of
the model. In particular, a more direct construction of $N$ commuting
conserved operators than the one recently given by Talstra and
Haldane \cite{TaHa94} would be highly desirable. Talstra and Haldane
exploited the fact that $H$ can be regarded at as the static limit of
the Sutherland Hamiltonian with spin degrees of freedom.

Note that the representation (\ref{10}) is a ``square root'' of
the Hamiltonian in the terminology of Shastry \cite{Shastry92}.
Introducing creation and annihilation operators of spin waves,
$\tilde{S}_k^+ := \tilde{S}_k^x - \i \, \tilde{S}_k^y$,
$\tilde{S}_k^- := (\tilde{S}_k^+)^+$, the spin wave spectrum is
easily rediscovered from (\ref{10}). To begin with, the
Hamiltonian reads
\beq
     H = \frac{2}{N} \sum_{k=1}^{N-1} k(k - N)
         (\tilde{S}_k^+ \tilde{S}_k^- + \tilde{S}_k^z
          \tilde{S}_{-k}^z + \tilde{S}_0^z - N/2) \qd.
\eeq
Now let $|0\>$ denote the ferromagnetic vacuum where all spins are
up. Then $\tilde{S}_k^- |0\> = 0$, $\tilde{S}_k^z |0\> = \de_{k0}
N/2 |0\>$, and thus $H|0\> = 0$. Since
\beq
     [H,\tilde{S}_k^+] = \frac{2}{N} \sum_{j=1}^{N-1} j(j - N)
                         (\tilde{S}_j^+ \tilde{S}_{k-j}^z -
                          \tilde{S}_{k+j}^+ \tilde{S}_{-j}^z -
                          \tilde{S}_{k-j}^+ \tilde{S}_j^z) \qd,
\eeq
it follows that
\beq
     H \tilde{S}_k^+ |0\> = k(k - N) \tilde{S}_k^+ |0\> \qd.
\eeq

In the su(2) representation $H$ commutes with the generators
\bea \label{13}
     J_0^{\a} & := & \tilde{S}_0^{\a} \qd, \\ \label{14}
     J_1^{\a} & := & \frac{1}{8} \e_{\a \be \g}
                     \sum_{j,k = 0 \atop j \ne k}^{N-1}
                     \ctg ((j - k) \p /N) \s_j^{\be} \s_k^{\g}
\eea
of a Yangian quantum group \cite{HHTBP92}. This fact explains the
high degeneracy of the spectrum of the Hamiltonian. With the aid
of the transformation (\ref{4}) $J_1$ takes the following form:
\beq
     J_1^{\a} = \i \, \e_{\a \be \g} \sum_{k=1}^{N-1}
                \left( \frac{N}{2} - k \right) \frac{1}{N}
                \tilde{S}_{-k}^{\be} \tilde{S}_k^{\g} \qd.
\eeq
This again may be considered as a ``square root'' of the expression
(\ref{14}) and complements the expression (\ref{10}) for the
Hamiltonian.

The su(2) Haldane-Shastry Hamiltonian and some of its
generalizations have been pointed out to be equivalent to
effective Ising models \cite{GeRu92,FrIn94}. That means that
they are equivalent to trivially interacting Fermion models,
where the Hamiltonian contains only particle number operators
$n_j = c_j^+ c_j$, as well. So far the
corresponding explicit transformations have been unknown. In
this letter we presented for the first time such a transformation
for the special case of the su(1$|$1) model.

F.G. gratefully acknowledges financial support by the Japan
Society for the Promotion of Science.

\end{document}